\documentclass[twocolumn,aps,tightenlines,floatfix,showpacs]{revtex4}
\usepackage{graphicx}
\newcommand{\be}{\begin{equation}}
\newcommand{\ee}{\end{equation}}

\begin{document}
\title{Many-Body Approximations in the sd-Shell Sandbox}

\author{R.A. Sen'kov$^{1,2}$, G.F. Bertsch$^{3}$, B.A. Brown$^1$,  Y.L. Luo$^3$,
and V.G. Zelevinsky$^1$}

\affiliation{$^1$National Superconducting Cyclotron Laboratory and
Department of Physics and Astronomy, Michigan State University,
East Lansing, MI 48824-1321, USA}
\affiliation{$^2$ Department of Physics, Novosibirsk State University,
Novosibirsk 630090, Russia}
\affiliation{$^3$Department of Physics and Institute of Nuclear Theory,
 Box 351560, University of Washington, Seattle, Washington 98915, USA}

\begin{abstract}
A new theoretical approach is presented that combines the
Hartree-Fock variational scheme with the exact solution of the
pairing problem in the finite orbital space. Using this
formulation in  the $sd$-space as
an example, we show that the exact pairing significantly improves
the results for the ground state energy.
\end{abstract}


\maketitle

\section{Introduction}

The classical Hartree-Fock (HF) approximation is a prototype of the
modern approach to the quantum many-body problem related to the
energy density functional \cite{KS65,dreizler90}. When applied to
complex nuclei, the density functional theory may provide a universal
description across the nuclear chart. The pairing interaction that
is present in nuclei as well as in fermionic condensed matter
systems is usually included in the Hartree-Fock-Bogoliubov (HFB)
form \cite{goodman79}. The well known deficiencies of the HFB
approach for mesoscopic systems follow mainly from its
non-conservation of particle number. As a result, unphysical
features are introduced into dynamics, the superfluid phase
transition appears too sharp, and the correlational energy produced
by pairing might be severely underestimated.  As was shown earlier
\cite{VBZ01,ZV03pairing}, the pairing part of the problem can be
solved numerically quite easily with the help of the seniority
representation in a spherical basis, and its exact solution
significantly improves the results.

It was also sketched in \cite{VBZ01} how other parts of the
interaction can be incorporated into the exact pairing method in the
approximate way that reminds the HF approach. This can be done in an
iterative fashion: the exact pairing solution using the starting
single-particle basis determines the actual occupation numbers;
these (in general, fractional) occupancies self-consistently
determine, in the HF spirit, an improved single-particle basis where
we again solve the pairing problem etc. until convergence. In this
way both mean-field features, deformation and pairing, are accounted
for. The main purpose of the current work is to further develop this
Hartree-Fock plus pairing correlation (HFP)
method that essentially is an intermediate step from the HF approach
towards the full shell-model (SM) description. On one hand, we would
want to keep the simplicity and modest computer demands as inherent
properties of the mean field approach. On the other hand, we take
into consideration pairing and other physical effects beyond the
simple HF, or mean field in general. We check our approach for the
$sd$-shell nuclei, where the SM with large-scale diagonalization
works perfectly \cite{sdsm} and can serve as a searchlight
illuminating the correct direction of motion. The success of this
attempt will allow the extension of the approach to heavy nuclei,
where the catastrophic growth of dimensions makes the complete shell
model solution unrealistic.

\section{Outline of the method}

As in most mean-field approaches, we formulate this method as
variational one. As in the shell model, we assume a general form of
the two-body Hamiltonian that includes the single-particle term $t$
and the (antisymmetrized) two-body interaction $V$:
\begin{equation}
\hat H =\sum_{ik} t_{i k} \, a^\dag_i a_k + \frac{1}{4}\,
\sum_{ijkl}V_{i j k l} \, a_i^\dag a_j^\dag a_k a_l . \label{1}
\end{equation}
The variational wave function $|\Psi\rangle$ will be defined below.
The wave function and all properties of the system follow from
the minimization of the expectation value
\begin{equation}
\langle \Psi |\hat H|\Psi\rangle.
\end{equation}
For our test of the methods, we will take for $V$ the USDB
interaction from the $sd$-shell model \cite{sdsm}. It will allow us
to compare the results obtained using our approximate method with
the exact shell model calculations in the same single-particle
model space.

The ground state wave function $|\Psi \rangle$ for a fixed particle
number $N$ can be presented as a superposition of basis states,
 \be
| \Psi \rangle = \sum_{d \in {\sl D}} C_d | d \rangle, \label{2}
 \ee
where each basis state $| d \rangle$ is a Slater determinant which
for $N$ fermions can be written as usual:
 \be
| d \rangle = a^\dag_{\nu_1} a^\dag_{\nu_2} \dots a^\dag_{\nu_N}| 0
\rangle.                                                \label{3}
 \ee
The single-particle states $\phi_\nu $ can be found with the help of
the variational principle as it is usually done in the HF method.
The approach is actually defined by the selection of the space $D$
spanned by the determinants $|d\rangle$. If we choose only one
Slater determinant as our variational wave function (\ref{2}), we
come to the standard HF approximation. If the manifold $D$ includes
all possible configurations, then we get the exact shell-model
solution.

In this article our choice is determined by the pairing phenomenon
which smears the Fermi surface and converts the Fermi-gas ground state
into a superposition of Slater determinants. In the case of a {\sl
spherical} system with the pairing forces taken as the $J=0, T=1$ part of
the two-body interaction (\ref{1}), we have seniority $s$ as a good
quantum number. For an even number of particles, the ground state
has $s=0$, while for an odd number $s=1$. In this simple case we can
construct the basis of Slater determinants $| d \rangle $ occupying
single-particle levels $|jm\rangle$ by pairs,
 \be
 \prod_{j; m>0} a_{j m}^\dag a_{j\tilde{m}}^\dag |0\rangle, \label{4}
 \ee
where $a_{j\tilde{m}}^\dag = (-1)^{j-m} a_{j -m}^\dag$ is the
creation operator for the time-conjugate single-particle state with
respect to $a_{j m}^\dag$. Here we omit all quantum numbers except
total angular momentum $j$ and its projection $m$.

The presence of other types of the interaction in general breaks
spherical symmetry and brings in the {\sl deformed} mean field. In
the case of a deformed nucleus, even if we had had only $J=0$ part
of the two-body interaction (\ref{1}) in the spherical
representation, we have to take into consideration a broader class
of pairs arising as a result of splitting and mixing of the original
spherical states by deformation. Here we limit ourselves by the case
of axially symmetric deformation, when the single-particle orbitals
$|\nu m\rangle$ are still characterized by the angular momentum
projection $m$ along with other quantum numbers $\nu$.

According to the Kramers theorem, the orbitals $|\nu m\rangle$ and
$|\nu -m\rangle$ are degenerate. However, the pairs may also be
formed by the states $m$ and $-m$ belonging to different sets of
remaining quantum numbers. Thus, for our basis Slater determinants
$| d \rangle$ we assume the following form:
 \be
\prod_{\nu, \kappa; m>0} a_{\nu m}^\dag a_{\kappa -m}^\dag
|0\rangle.                                         \label{5}
 \ee
We construct the variational wave function (\ref{2}) as a
superposition of the Slater determinants (\ref{5}) for a given
particle number. Using such a form we hope to correctly account for
pairing correlations in the deformed case at the same time crucially
reducing the dimension of space $D$ in comparison with the full
shell model. Actually prescriptions (\ref{4},\ref{5}) are valid only
for an even number of particles. For an odd particle number, we use
the same eq. (\ref{5}) but add one additional creation operator that
corresponds to the odd particle. The odd particle can be placed in
any empty single-particle state, and the states are divided into
classes with a definite value of the angular momentum projection.

In the current application of our method we make a simplifying
approximation treating protons and neutrons separately. It means
that, though we consider the full two-body interaction (\ref{1})
including $T_z=0$ part, the variational function (\ref{2}) is
constructed as the product of proton and neutron parts. Clearly we
are losing here proton-neutron correlations although their mutual
contributions to the mean field are fully accounted for.

The variation over amplitudes $C_d$ with the additional
normalization condition of the wave function, $\langle \Psi | \Psi
\rangle = 1$, leads us to the usual set of equations,
 \be
\sum_{d'} \langle d | \hat H | d' \rangle C_{d'} = E C_d. \label{6}
 \ee
The matrix elements $\langle d| \hat H | d' \rangle $ are calculated
for the determinants built on a given single-particle basis, and
equations (\ref{6}) are solved numerically. The mean-field basis is
found from the self-consistent HF equations:
 \be
 h(\rho)\phi_{\nu} = \epsilon_\nu \phi_\nu,              \label{7}
 \ee
where
\begin{equation}
h(\rho)= t + V(\rho)                                   \label{8}
\end{equation}
is the single-particle HF Hamiltonian, $ \epsilon_\nu$ are the
single-particle energies, and $\rho$ is the density matrix
self-consistently determined by
 \be
\rho_{i j} = \langle \Psi | a^\dag_j a_i | \Psi \rangle. \label{9}
 \ee
The mean field potential is given by its matrix elements,
\begin{equation}
V_{ij}(\rho)=\sum_{kl}V_{iklj}\rho_{lk}.             \label{10}
\end{equation}
In this conventional mean field formulation, the potential
(\ref{10}) contains the direct and exchange contributions. The
pairing effects, with strict particle number conservation, are
contained in the superposition of the Slater determinants (2) used
instead of the single HF determinant. The whole construction can be
further improved by using the exact but more complicated variational
approach relating the single-particle basis to the full set of the
coefficients of the superposition (2).  Such a possibility will be
considered in the future.

The HFP scheme of solution is the following:
\begin{itemize}
\item Start with the spherical single-particle basis $| \kappa m\rangle$
\item Choose in this basis the initial diagonal density matrix $\rho$
corresponding to occupation numbers specific for prolate or oblate
shapes (pairs with small or large $|m|$, respectively)
\item Solve the HF variational equation (\ref{7}) and get the
single-particle spectrum ($\phi_\nu, \epsilon_\nu$), in general
corresponding to a deformed field
\item Construct the ``paired" class of many-body basis wave functions
according to eq. (\ref{5}) and calculate the matrix elements of the
Hamiltonian $H$
\item Solve the variational equation (\ref{6}) and obtain the ground state
wave function
\item Calculate the next-step density matrix (\ref{9})
\item Repeat the procedure starting from the step three until convergence
\end{itemize}

The converged results will certainly be a local minimum of eq. (2).
Exploration of different starting choices in step 2 is needed to
find a global minimum.  In our study here we start with a spherical
single-particle basis (because the USDB interaction is so defined) but
in principle any convenient axial basis could be used. In the end, the ground 
state energy can be found as the Hamiltonian expectation value over the 
resulting ground state wave function $|\Psi \rangle$,
\be E_{{\rm HFP}} = \langle \Psi | \hat H | \Psi \rangle.          \label{11}
\ee
\begin{figure}
\scalebox{0.5}{\includegraphics{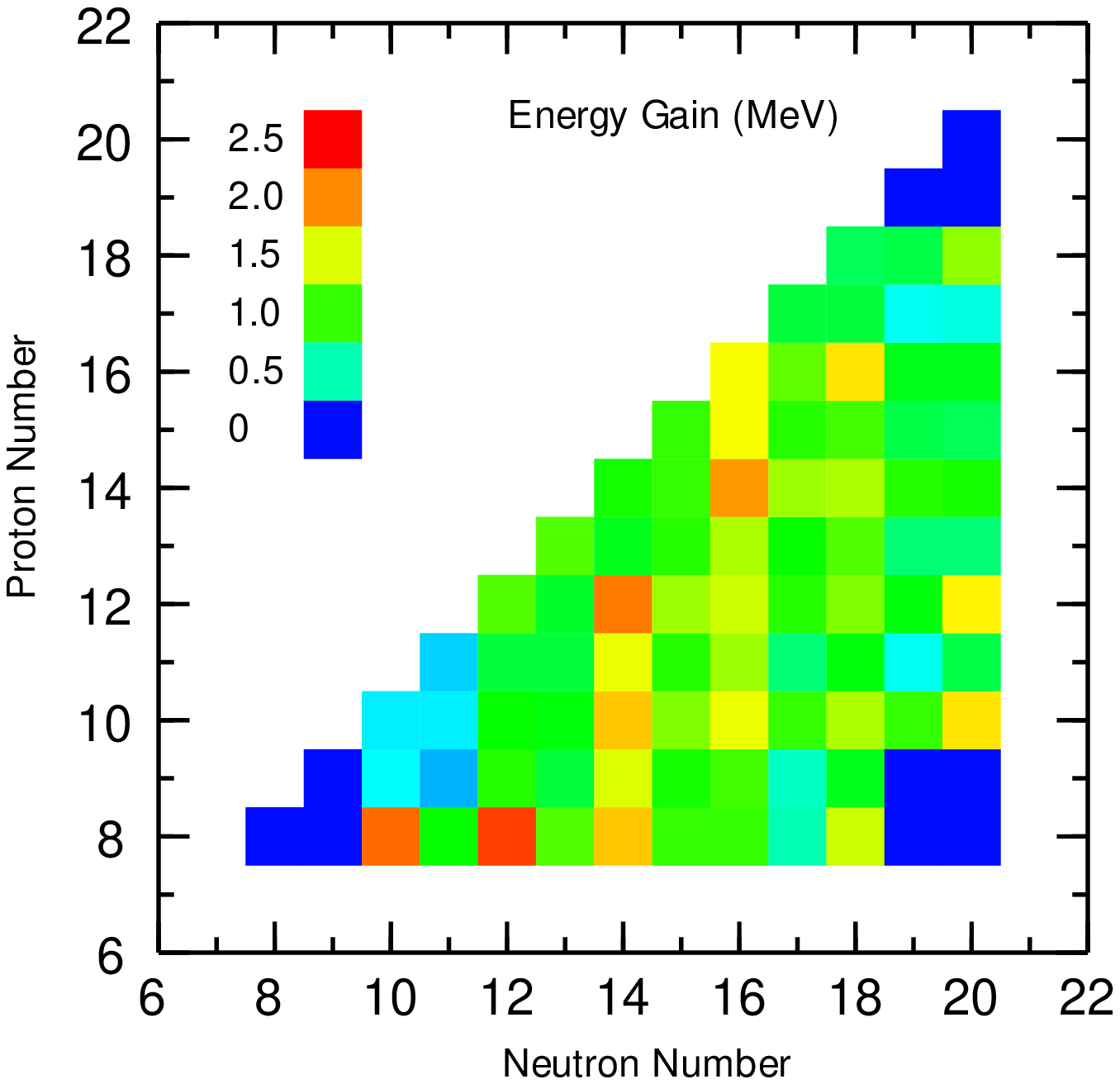}}
\caption{Pairing correlation energy eq. (\ref{12gb}) for all
$sd$-shell nuclei.}
\scalebox{0.3}{\includegraphics{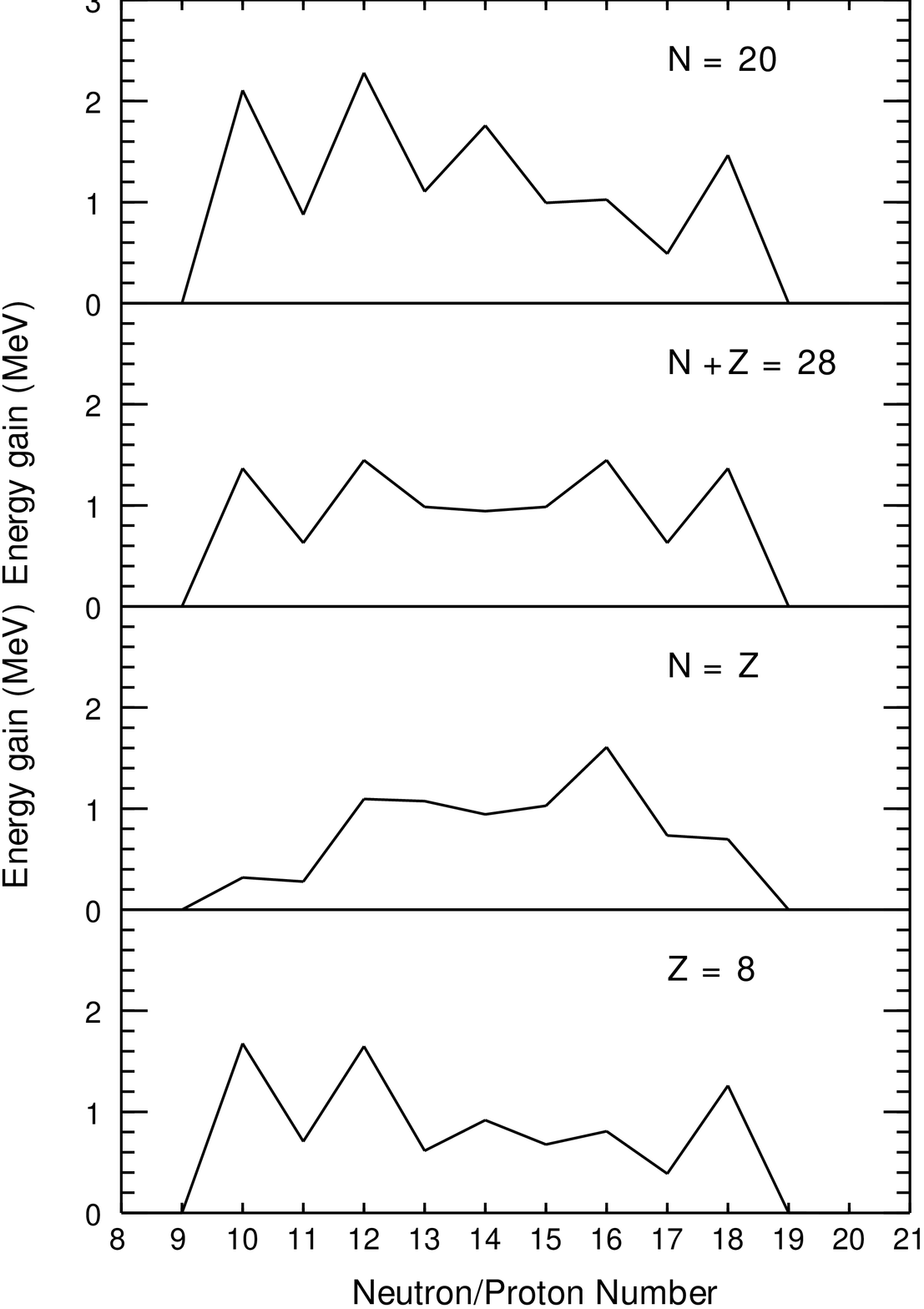}}
\caption{Four selected cuts on the correlation energies displayed in
Fig. 1 }
\end{figure}

\begin{figure}
\scalebox{0.5}{\includegraphics{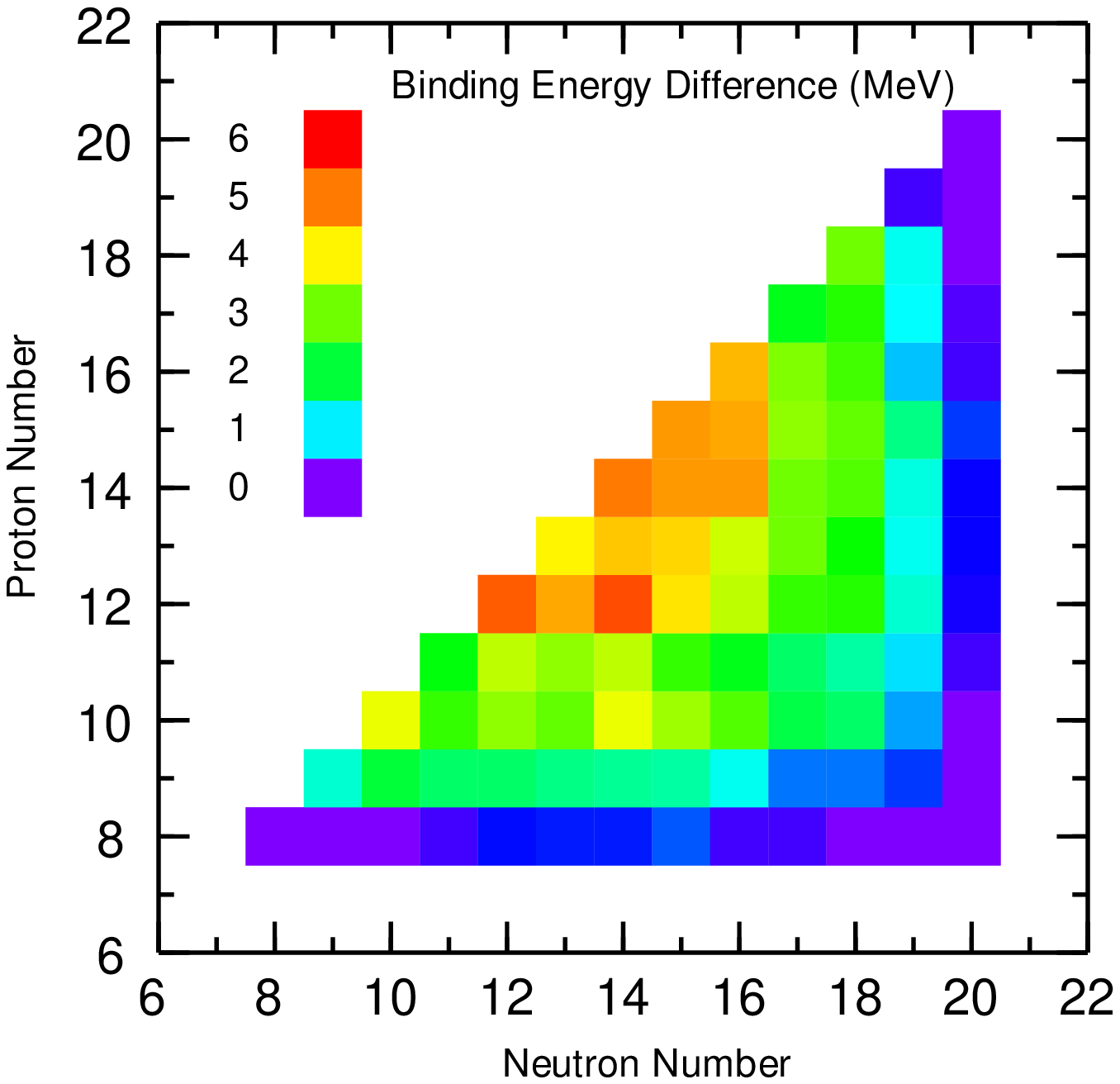}} \caption{Energy difference
between the exact results and our HFP model for all $sd$-shell
nuclei.} \scalebox{0.3}{\includegraphics{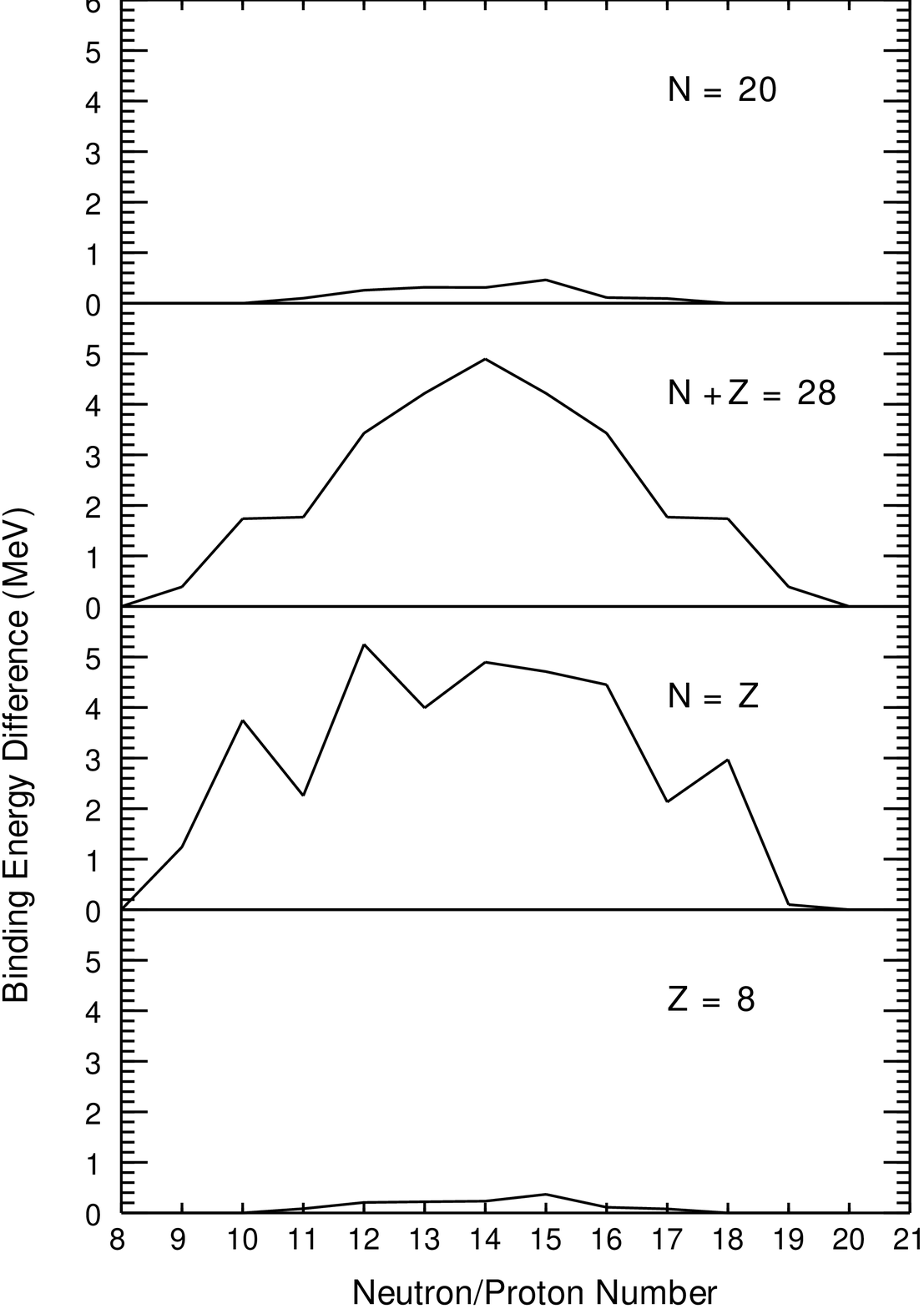}} \caption{Energy
difference between the exact results and our HFP model for four
selected cuts through the $sd$ shell.}
\end{figure}

\section{Results}

We performed calculations of ground state energies for all nuclei
within the $sd$-shell region, from $^{17}$O to $^{39}$Ca.
Our results \cite{table} are summarized in Figs. 1-5. In Figs. 1 and
2 we show the energy gain from HFP compared to HF,
\begin{equation}
E_{\rm corr} = E_{\rm HF}   - E_{\rm HFP}. \label{12gb}
\end{equation}
Typical values are one to a few MeV.  One observes the well-known
odd-even staggering that is characteristic of pairing. In the
conventional HFB approach the pairing correlation is zero for many
of these nuclei, including cases such as $^{24}$O where the
spherical shell gap is too large, and cases such as $^{20}$Ne,
$^{24}$Mg and $^{28}$Si where the deformed shell gap is too large to
support BCS type pairing. The HFP method gives some correlation
energy for all $sd$-shell nuclei for which there are at least two
active particles ( $9 < N < 19$ or $9 < Z < 19$ ).
In the practical solution of the equations we find that many
$sd$-shell nuclei have two or three energy minima. To have some
confidence that we have found the lowest energy solution we start
with several initial values of the density matrix including those
that are prolate and oblate deformed, spherical and random.

In Figs. 3 and 4 we show the difference between the HFP result and
the exact shell-model energy both obtained with the USDB
Hamiltonian. For comparison of the methods, the full solution for
the ground state of $^{28}$Si must take into account 93,710
$M$-scheme Slater determinants. When projected onto good $J=0$ there
are 9,216 states, and when projected onto good $J=0, T=0$ there are
839 states. The HFP method requires 92 determinants for protons and
92 for neutrons. The difference is clearly peaked at the $N=Z$ line
that can be explained by the proton-neutron pairing being not
accounted for in the calculations. 

The HFP solution is very close to the exact solution around the
edges of the $sd$ shell (see Fig. 4). These nuclei are spherical and
the HFP method is equivalent to the spherical exact-pairing method
discussed in \cite{VBZ01,ZV03pairing}. The largest deviation from
exact is for nuclei near the middle of the $sd$ shell. There are
still pairing contributions for deformed nuclei, but the result is
different from the naive expectation of just adding "spherical"
contributions. For example, as shown in Fig. 2, the correlation
energy is only about 400 keV for the deformed $^{20}$Ne, compared to
a total of about 3.4 MeV that would be obtained just from adding the
1.7 MeV correlation energies obtained for two neutrons and two
protons in a spherical basis (e.g. $^{18}$O and $^{18}$Ne).

Finally in Fig. 5 we show the intrinsic quadrupole moment obtained
for the lowest energy solutions for all $sd$-shell nuclei. One
observes the well known region of strongly prolate nuclei near
$^{24}$Mg. $^{28}$Si is the most strongly oblately deformed, and
there is an island of weak oblate deformation around $^{31}$Si. It
would be interesting to use our $sd$-shell sandbox to clarify the
general question of why most nuclei are prolate deformed
\cite{tajima}, by exploring the HFP results with different (but
realistic) Hamiltonians.

\begin{figure}[h]
\scalebox{0.5}{\includegraphics{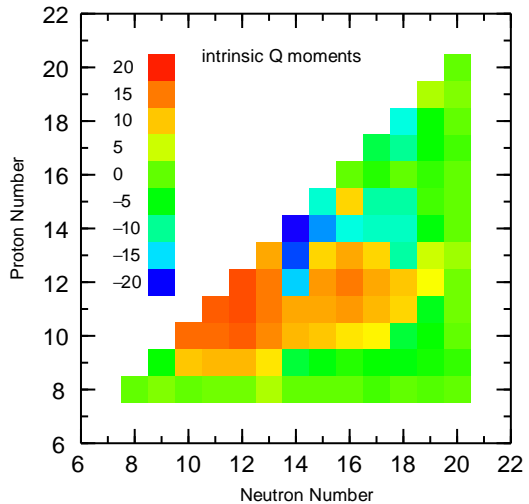}} \caption{Intrinsic
quadrupole moment $Q$, in units $b^2$ ($b$ is the oscillator parameter),
for the lowest energy configuration for all $sd$-shell
nuclei.}
\end{figure}
\section{Conclusion}

Obviously, the HFP is still far from adequate away from semi-magic
nuclei.  The angular momentum non-conservation is certainly a
significant deficiency of the wave function, that when repaired will
introduce additional correlation energy.  There are a number of ways
that rotational correlation energies can be calculated, and we are
optimistic that HFP wave functions can be used as a better starting
point.

Besides rotational energies,
proton-neutron pairing effects are omitted in our wave
functions. As known \cite{martinez99}, such pairing correlations are
quite strong close to the $N=Z$ line; they should be included at the
next stage of development.
Effective
Hamiltonians for the HFB solution are explored in \cite{rod}.
Finally, some improvement may follow from including the non-axial
configurations with the pairing between more general time-conjugate
orbitals (most probably, the mean field in $^{24}$Mg is triaxial
\cite{cole89},\cite{bender}).

\section{Acknowledgements}
Supported by UNDEF-SciDAC  DOE grant   DE-FC02-07ER41457
and NSF grants PHY-0555366 and PHY-0758099.

\end{document}